\documentclass{IEEEtran}
\usepackage{framed,multirow}
\usepackage{cite}
\usepackage{amsmath,amssymb,amsfonts}
\usepackage{algorithmic}
\usepackage{graphicx}
\usepackage{textcomp}
\usepackage{multirow}
\usepackage{comment}
\usepackage{tabu}
\usepackage{makecell}
\usepackage[colorlinks=true]{hyperref}

\newcommand{\rv}[1]{\textcolor{black}{#1}}

\begin{document}
\title{Follow My Eye: Using Gaze to Supervise Computer-Aided Diagnosis}
\author{Sheng Wang, Xi Ouyang, Tianming Liu, Qian Wang*, Dinggang Shen*, \IEEEmembership{Fellow, IEEE}
\thanks{Sheng Wang and Xi Ouyang are with the School of Biomedical Engineering, Shanghai Jiao Tong University, Shanghai, China. (e-mail: \{wsheng, xi.ouyang\}@sjtu.edu.cn).}
\thanks{Tianming Liu is with the Department of Computer Science, University of Georgia, GA, USA. (e-mail: tliu@cs.uga.edu).}
\thanks{Dinggang Shen and Qian Wang are with the School of Biomedical Engineering, ShanghaiTech University, Shanghai, China. (e-mail:\{dgshen, wangqian2\}@shanghaitech.edu.cn). Dinggang Shen is also with the Department of Research and Development, Shanghai United Imaging Intelligence Co., Ltd., Shanghai, China. }
\thanks{This work was supported in part by Science and Technology Commission of Shanghai Municipality (19QC1400600 and 21010502600), National Natural Science Foundation of China (62131015), and The Key R\&D Program of Guangdong Province, China (2021B0101420006).}
}

\maketitle

\begin{abstract}
When deep neural network (DNN) was first introduced to the medical image analysis community, researchers were impressed by its performance. However, it is evident now that a large number of manually labeled data is often a must to train a properly functioning DNN. This demand for supervision data and labels is a major bottleneck in current medical image analysis, since collecting a large number of annotations from experienced experts can be time-consuming and expensive.
In this paper, we demonstrate that the eye movement of radiologists reading medical images can be a new form of supervision to train the DNN-based computer\rv{-}aided diagnosis (CAD) system. Particularly, we record the tracks of the radiologists’ gaze when they are reading images. The gaze information is processed and then used to supervise the DNN’s attention via an Attention Consistency module. To the best of our knowledge, the above pipeline is among the earliest efforts to leverage expert eye movement for deep-learning-based CAD. 
We have conducted extensive experiments on knee X-ray images for osteoarthritis assessment. The results show that our method can achieve considerable improvement in diagnosis performance, with the help of gaze supervision. 
\end{abstract}


\section{Introduction}
\label{sec:introduction}

\IEEEPARstart{M}{edical} image analysis plays an essential role in modern medical practice. One of the most popular trends in this area is to apply deep learning techniques \rv{to} computer\rv{-}aided diagnosis (CAD), which has achieved great \rv{success} in the past decade. The progress should be attributed to not only novel deep learning techniques but also large volumes of carefully annotated data. 
However, annotating a large medical image dataset is usually expensive and time-consuming, which requires experienced clinical experts to examine the images, fuse them with other examinations and lab reports, and sometimes even consult with more experts.

\begin{figure}[t]
    \centering
    \includegraphics[width=0.48\textwidth]{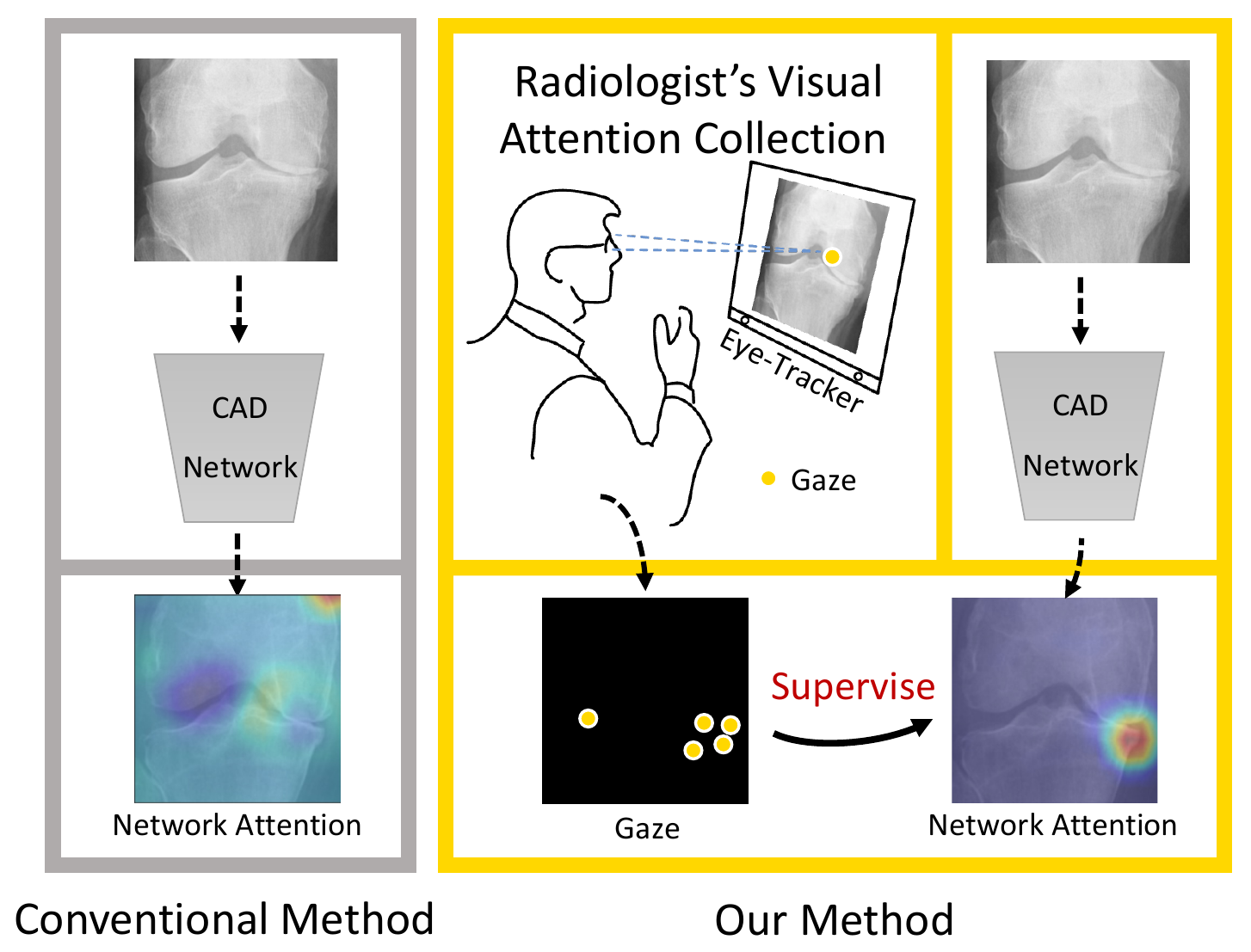}
    \caption{Our proposed method (in the right panel) makes use of the human gaze to train the network, compared to the conventional method (in the left panel). The supervision upon the network’s attention can optimize the network’s performance in classification and abnormality localization.}
    \label{fig:cover_image}
\end{figure}
When the deep learning models are trained with only image-level annotations (e.g., disease labels of the subjects), it is often difficult for the models to attain abnormality localization, which then weakens the interpretability of the models used in CAD. Therefore, integrating multi-level supervision has become a hot topic in recent years. In addition to exploring the dataset by image-level annotations, the deep networks can benefit from (a small number of) extra annotations of finer granularity. For example, Li et al.~\cite{li2018tell} proposed a guided attention inference network to utilize both the image-level and pixel-level annotations on natural images. Ouyang et al.~\cite{ouyang2020learning} presented a chest X-ray diagnosis framework, by using bounding boxes for rough abnormality annotations, \rv{improving} both diagnosis and abnormality localization simultaneously. These works prove that extra fine-level supervision can improve accuracy in CAD, as well as the interpretability and robustness of the trained deep models.

Although a small amount of extra fine-level annotations is relatively easy to collect, the process can still be prohibitively expensive in practice. \rv{Particularly, for each clinical center, imaging device, or even clinician, one may need to repeat the certain fine-level annotating process to tune the pre-trained deep networks.} It is thus a desire that the annotating process should incur little impact on the daily clinical workflow. To this end, we propose a solution to utilize the expert's gaze information to better model the diagnosis network in this paper. The gaze information is collected by eye-trackers, without interrupting the normal clinical workflow of the radiologists when they read the images. Then, the gaze information is processed and used as additional supervision to train the deep network for CAD.

We note that tracking eye movement is not completely new to the field of radiology. In 1981, Carmody et al. \cite{carmody1981finding} found that the eye scanning strategy of the radiologist impacted the false negative errors when looking for nodules from chest X-ray\rv{s}. More recently, it is found that the visual patterns are associated with diagnostic performance for lesion detection when reading mammography~\cite{kundel2008using,voisin2013investigating} and CT images~\cite{bertram2016eye,mallett2014tracking}. The literature studies have implied that the potentials of the radiologist’s gaze data can be high in improving disease diagnosis~\cite{wu2019eye,brunye2019review}.

In this paper, we propose a gaze-guided attention network for automatic disease diagnosis. 
\rv{It is among the pilot studies to use radiologist’s gaze to improve diagnosis performance for deep-learning-based CAD systems.} 
We utilize an online class activation map (CAM) strategy \cite{ouyang2020learning} to calculate the attention map, which represents the evidence for network decisions. At the same time, the gaze maps are acquired to reflect human attention for disease regions when the radiologists read the images. \rv{In order to make the network mimic the radiologist's decision-making process, we propose an attention consistency module to regularize the network attention maps so that they are directly regularized to human attention.}

Finally, we conduct comprehensive experiments on a publicly available knee X-ray dataset~\cite{chen2019fully}. 
The experimental results prove the value of radiologists' gaze and the effectiveness of our proposed method.

\rv{We summarize our major contributions as follows}.
\begin{itemize}
\item \rv{This is among the \rv{pilot studies} to demonstrate the critical role of radiologist's eye movement in a deep-learning-based CAD system.} We implement a simple yet effective deep learning solution for utilizing the guidance from the radiologist’s gaze. We demonstrate that the extra supervision from expert gaze can improve the accuracy, robustness and interpretability of the CAD system.

\item We present an effective way to collect and process the radiologist’s gaze track. As extra supervision to deep learning, the collection of the gaze data is conducted by the eye-trackers under the screen, with little impact on the radiologist’s routine workflow. The data collection and processing scheme is convenient and economically efficient.

\item We develop a user-friendly software including the eye-tracker Python API, GUI and toolkit for image reading, gaze collection, and post-processing\footnote{Codes will be available at \href{https://github.com/JamesQFreeman/MICEYE}{https://github.com/JamesQFreeman/MICEYE}.}. Our work will potentially benefit other researchers and the entire CAD community.
\end{itemize}

\section{Background}
In this section, we first review the attention mechanism and attention guidance in computer vision and deep learning. Then, we introduce the related works regarding human visual attention, radiologist’s gaze, and image-based intelligent diagnosis.

\subsection{Attention Mechanism in Deep Networks}
The attention mechanism has been widely investigated in the deep learning framework recently, which has shown promising results in computer vision\cite{hu2018squeeze,woo2018cbam,fu2019dual} and natural language processing.
\rv{We categorize the attention-based convolutional neural networks (CNN) into two major classes, including \textit{channel attention} and \textit{spatial attention}.}

In channel attention, different convolutional channels in the network are assumed to have different impacts in generating the network output. Meanwhile, in spatial attention, different spatial locations are assumed to contribute differently to the output, resulting in various saliency assigned to individual spatial parts of the input images. A common approach to reveal the intrinsic attention of CNN is to examine the gradient-based activation in back-propagation, such as using CAM~\cite{zhou2016learning} and Grad-CAM~\cite{selvaraju2017grad}.

\rv{CAM is widely used in computer vision to look into the working mechanism of the network. Other than visualization, it is also used in the optimization to give more control during the training process.} Based on CAM, Li et al.\cite{li2018tell} proposed a framework to guide the network attention to make the attention map more accurate and descriptive. They demonstrated that constraining the network attention with a segmentation mask could improve the performance of the network in the weakly-supervised image segmentation task. Ouyang et al. \cite{ouyang2020learning} presented a solution to constrain the network attention with bounding boxes for abnormality. They improved the accuracy in chest X-ray diagnosis and achieved a better capability of abnormality for localization.

There are also many works that make use of human visual attention for spatial attention in various computer vision tasks. Karessli et al.\cite{karessli2017gaze} used gaze for zero-shot image classification. Li et al.\cite{li2014secrets} used gaze to improve the object segmentation performance. Lin et al.\cite{lin2020interactive} used the mouse click as a critical human attention source in interactive segmentation and achieved significantly better performance.

To our best knowledge, there are few research studies leveraging human visual attention in learning-based medical image analysis. A related contribution was reported in Li et al.~\cite{li2019attention}, where they collected a database including retina fundus images, diagnosis labels and respective mouse click maps for glaucoma detection. Based on the mouse clicks, they proposed an attention-based CNN to detect glaucoma in fundus images. They showed the effectiveness of utilizing human attention. However, the attention in mouse clicks still needs additional human effort, which is in sharp contrast to eye-tracking data \rv{that make} minimal impact on radiologists.

\subsection{Eye-Tracking Research in Radiology}

Visual attention has been proven as a useful tool to understand and interpret radiologist's clinical decisions. In 1981, Carmody et al.~\cite{carmody1981finding} published one of the first eye-tracking studies in radiology, where they studied the detection of lung nodules in chest X-ray films. Four radiologists participated and their eye movements were recorded using special glasses based on the corneal reflection technique. The participating radiologists were instructed to press a key when they found a nodule in the X-ray film. The study found that the false-negative errors in reading the X-ray images could be impacted by the eye scanning strategies used by individual radiologists.

Eye-tracking studies are also conducted on other specialties such as mammography. \rv{Lago et al.~\cite{lago2020measurement} demonstrated that the radiologist's field of view can be very different across different imaging modalities and targets.} Kundel et al.\cite{kundel2008using} gathered eye-tracking data and found that 57\% of cancer lesions were located within the first second of viewing. Voisin et al.\cite{voisin2013investigating} investigated the association between gaze patterns and diagnostic performance for lesion detection in mammograms. It is found that gaze fixations are highly correlated with radiologists’ diagnostic errors. 

There are also studies that focus on CT and MRI. Bertram et al.\cite{bertram2016eye} investigated the image markers of visual expertise using abdominal CT images where the eye-tracking data showed that specialists react with longer fixations and shorter saccades when encountered with the presence of lesions. Mallett et al.\cite{mallett2014tracking} focused on CT colonography videos, which were interpreted by 27 experienced radiologists and 38 inexperienced radiologists. The eye-tracking data indicated that the experienced readers had higher rates of polyp identification than inexperienced readers as evidenced by multiple pursuits when examining polyps. Stember et al.\cite{stember2020integrating} used eye-tracking data to label brain tumor MRI scans. More related studies are surveyed in \cite{brunye2019review, wu2019eye}.

\subsection{Eye-tracking in CAD}
\rv{
Recently, many studies investigate eye-tracking in medical image analysis from the deep learning perspective. 
In~\cite{stember2019eye}, Stember et al. used eye-tracking data and speech recognition to automatically label tumors in brain images, which is faster and easier than the conventional manual annotation using mouse clicking and dragging.
Wen et al. \cite{wen2016computational} offered a computational assessment of visual search strategies in CT images.
Mall et al.~\cite{mall2019missed} investigated the relationship between human visual attention and CNN in finding missing cancer in mammography. They also modeled visual search behavior of radiologists for breast cancer using CNN~\cite{mall2018modeling}.}

\rv{Among all published works, Karargyris et al.~\cite{karargyris2021creation,karargyris2020eye} are relatively close to our work. They developed a dataset of chest X-ray images, gaze coordinates, and reports in text. Then they demonstrated to incorporate the effectiveness and richness of eye-tracking information from the gaze data. There were two forms to leverage the gaze data, i.e., ``temporal heatmaps" and ``static heatmaps". In the ``temporal heatmaps", the gaze passed the convolutional encoder and then concatenated with the information from the image. In the ``static heatmaps", a multi-task framework performed classification for diagnosis and predicted eye gaze at the same time.}

\begin{figure}[t]
    \centering
    \includegraphics[width=0.48\textwidth]{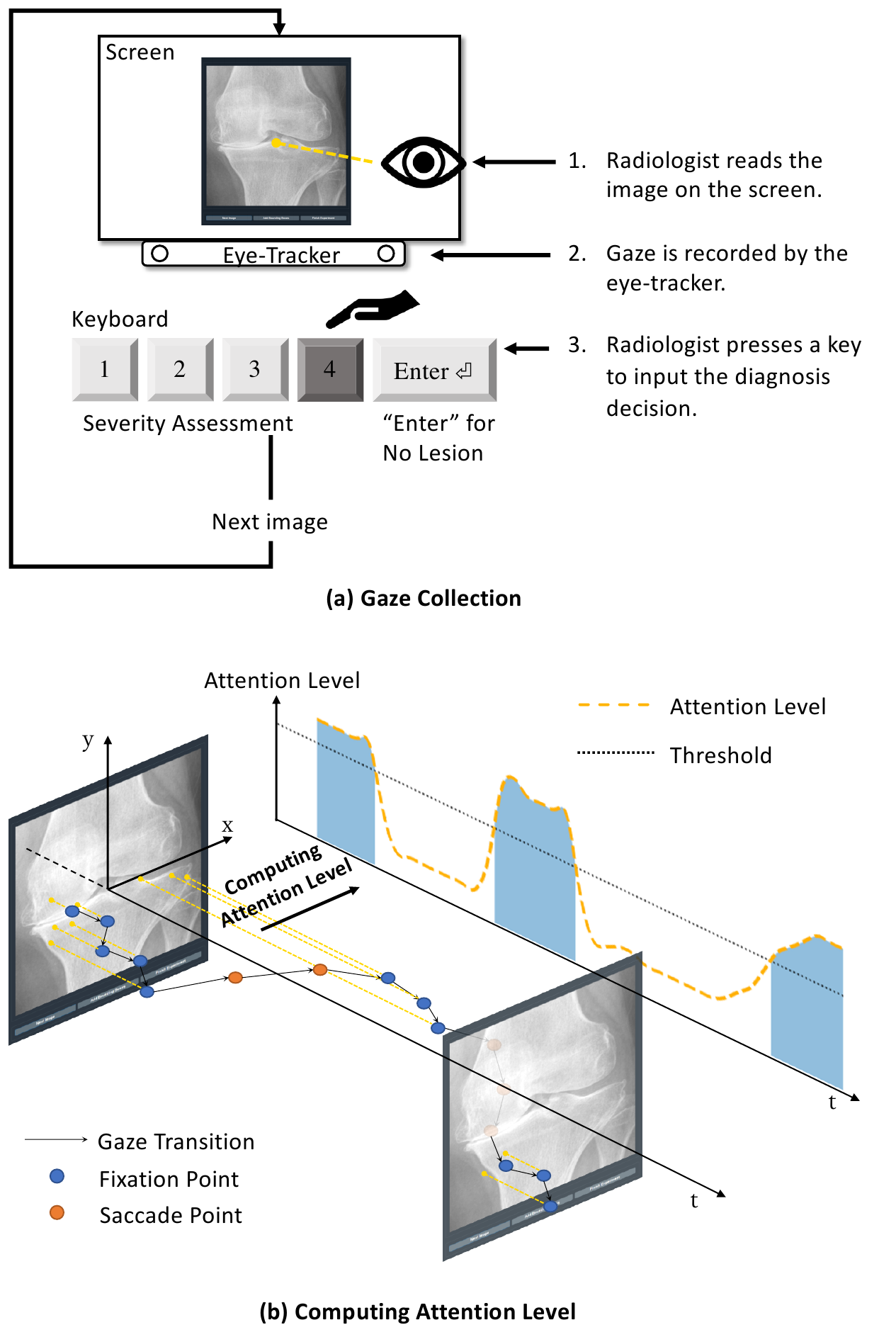}
    \caption{(a) Radiologists first read a knee X-ray image until they make a diagnosis decision. Their eye movement is recorded by the eye-tracker. The decision is recorded by the key pressed: ``1-4" represent KL Grade 1 - KL Grade 4, and ``Enter" is ``normal". (b) The collected gaze is a 3-dimensional tuple: $x,y$ are spatial coordinates and $t$ is a timestamp. Then the gaze points are used to compute the attention level. Then gaze points are further separated into the saccade points and the fixation points, respectively. Note that the saccade points mark the fast movement of the eyes while the fixation points mark the focusing locations of eyes.}
    \label{fig:eye movement}
\end{figure}
\section{Method}

\begin{figure*}[h]
    \centering
    \includegraphics[width=1\textwidth]{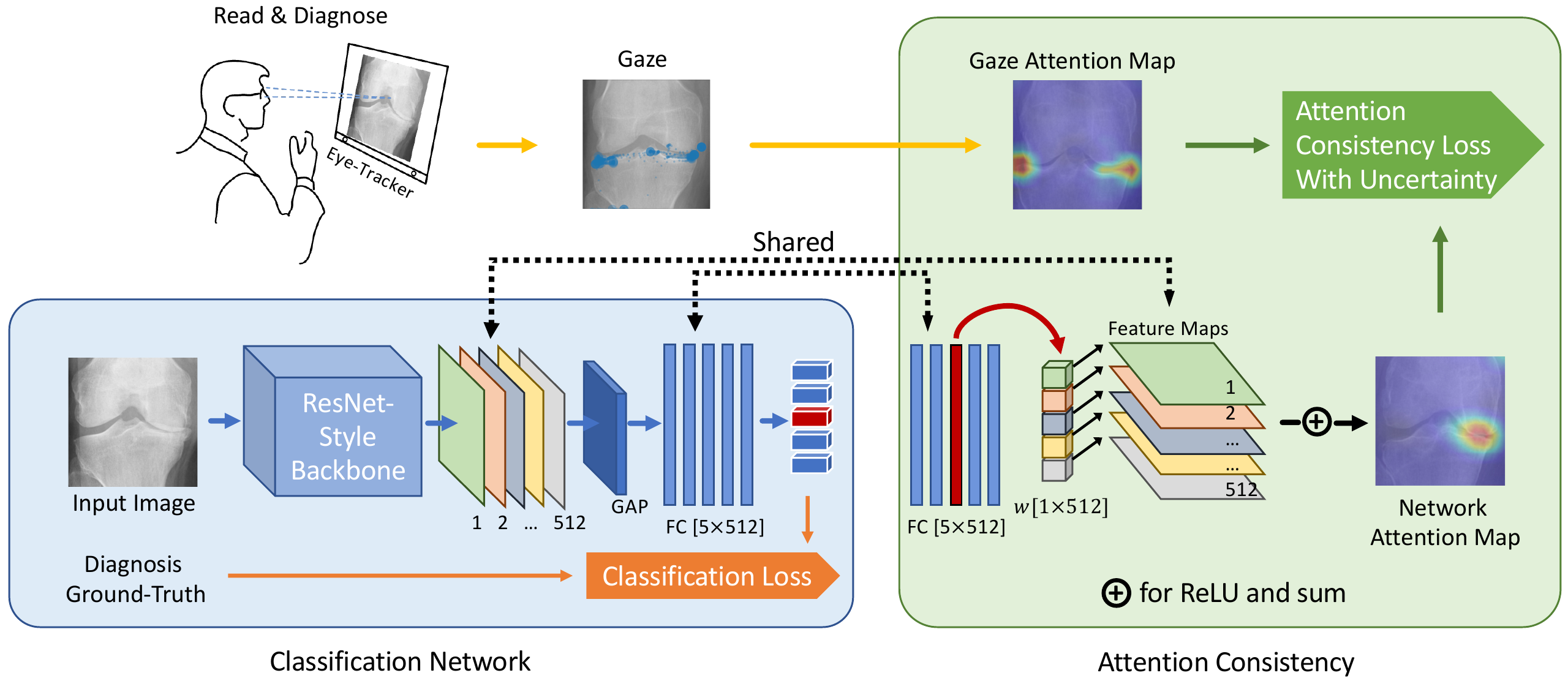}
    \caption{The proposed GA-Net consists of two parts, i.e., the classification network, and the attention consistency (AC) module. Both parts are activated during the training, yet only the classification network is used for inference.}
    \label{fig:framework}
\end{figure*}

The overall framework of this research consists of two main parts: 1) gaze collection and post-processing, and 2) modeling of the gaze-guided attention network (GA-Net). To access the gaze information from the radiologists, we develop a unified system for both gaze collection and post-processing, tailored to X-ray images. We use this system for collecting gaze information for osteoarthritis (OA) assessment, which is one of the most common joint diseases in the knees. \rv{Two musculoskeletal radiologists (with 2 and 5 years of clinical experience) took part in our experiment.} 

The disease of OA usually appears a lack of articular cartilage integrity as well as prevalent changes associated with the underlying articular structures, and may develop into joint necrosis or even disability if there is no early intervention \cite{glyn2015osteoarthritis}. We aim to utilize the gaze information of the radiologist when reading the knee X-ray images, to help improve the CAD system for OA assessment.
Therefore, we propose a gaze-guided attention network (GA-Net) for this task with the attention consistency to ensure that the network can focus on the disease regions like the radiologists. Since the collected gaze information contains high inter-observer variability (i.e., due to different reading habits of different doctors as well as \rv{other external conditions}), we propose to use homoscedastic uncertainty to quantify gaze maps.
In the subsequent sections, we will \rv{introduce} each part of our proposed framework in detail.

\subsection{Gaze Collection and Post-processing}
\label{gaze collection and post-processing}
The high-quality gaze maps are essential since it is the supervision information for the training of the following CAD system. To this end, we will first introduce the details of gaze collection and processing in the following.

\begin{figure*}[t]
    \centering
    \includegraphics[width=0.93\textwidth]{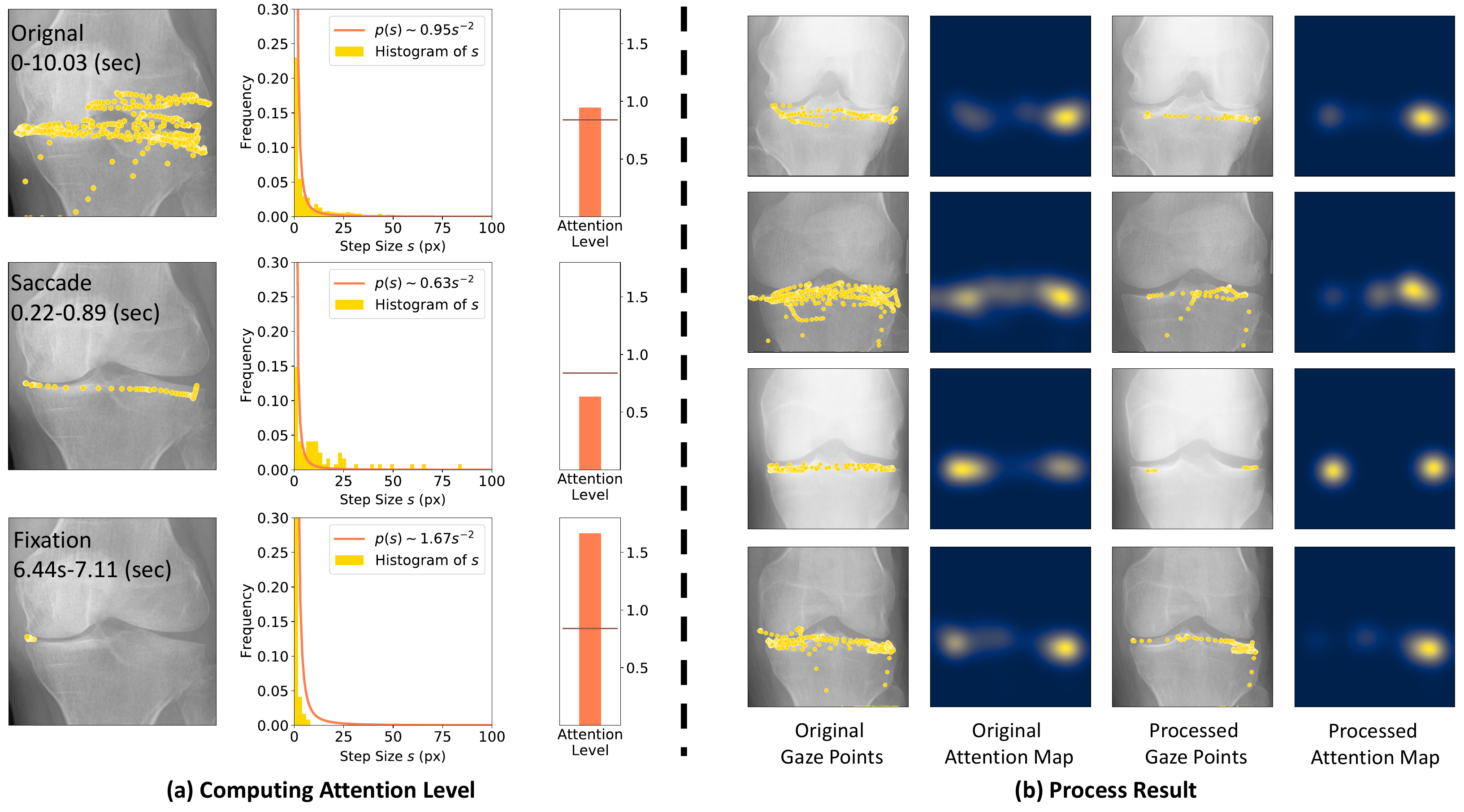}
    \caption{\textbf{(a)} Demonstration of computing the attention level from the tracks of the gaze points. The first column shows the scattering of the raw gaze points. The second column shows the histograms of gaze movement's step length $s$, respectively. The third column shows attention levels calculated by each histogram. \textbf{(b)} Gaze points and the corresponding attention maps before and after processing. The first left column shows the original eye movement tracks of four example images, and the third column shows the processed tracks. The second and the fourth columns show the corresponding gaze attention maps before and after processing, respectively.}
    \label{fig:denoise-comparison}
\end{figure*}

\textbf{Gaze Collection.} We collect the eye-tracking data with the Tobii 4C remote eye-tracker that records binocular gaze data at 90 Hz. We implement a customized data collection software in Python using the manufacturer-provided SDK. The software is made publicly available with this paper. Our software logs the reader’s on-screen gaze locations with the corresponding timestamps. \rv{The readers are seated in front of a $1920 \times 1080$ 27-inch LCD screen, to simulate the clinical working condition. In our experiment, the distance from the screen to the volunteers' eyes is about $50cm$. The images are displayed at $800\times 800$, which are $25\times 25 cm^2$ on the screen. Our eye-tracking primarily records the narrow, high-resolution central vision~\cite{carrasco2011visual}. More details about the setup and discussions on different types of visual attention can be found in the supplementary material.}

Our data collection paradigm is illustrated in Fig. \ref{fig:eye movement}(a). Note that we make the demonstration by reading knee X-ray images, for the purpose of OA assessment. \rv{First, radiologists enter their basic user profiles such as the name, gender and age, followed by a standard 5-point calibration routine~\cite{karessli2017gaze}.} Next, the radiologist follows a cycle of two steps, namely reading and diagnosis, to complete the diagnosis task on an X-ray image. During the reading step, an image is randomly drawn from our training dataset and shown to the radiologist, as the radiologist reads the image until they feel confident to reach a diagnosis decision. In the diagnosis step, the radiologist types in the decision by pressing the number keys in the keyboard, e.g., ``1-4” used for representing KL-Grade 1 - KL-Grade 4, and ``Enter"  for normal. We record the eye movement during the entire cycle of reading and diagnosis steps. We have a rest of 2 minutes for every 20 images to reduce radiologists’ fatigue.

\textbf{Attention Level.} \rv{In our experiments, we notice that the raw data from the eye-tracker contains information that is irrelevant to our task such as covert attention~\cite{carrasco2011visual} and inter-observer variability. In addition, during the collection, we have found the radiologist may not always fully concentrate on looking at the screen.} Therefore, we must find a way to identify the segments of eye movement that really focus on the lesions in the images.

As shown in Fig. \ref{fig:eye movement}(b), it can be observed that eye movement recording mainly contains the saccade (fast eye movement~\cite{liversedge2011oxford}) and fixation points in our experiment.
While the fixation points indicate the locations that the radiologist's eyes are focusing on, the saccade points contain little attention information related to the disease regions. Some simple criteria such as duration and velocity thresholds, however, do not perform well since readers have distinct habits and different distances to the screen.

To address this issue, we introduce a hypothesis to model the visual scan paths and denoise the gaze points. Brockmann et al.~\cite{brockmann1999human} suggested human eye scan paths are geometrically similar to a prominent class of random walk known as Levy flight, in which the length of each walking step is subject to a heavy-tailed distribution. 
The step length refers to the moving distance between two adjacent gaze points sampled by the eye-tracker.

Here, following~\cite{liu2013semantically}, we use power distribution to model the gaze step. Let $s$ be the step length which is subject to the following distribution
\begin{equation} \label{levy-flight}
p(s) = \gamma s^{-2},
\end{equation}
where $\gamma$ controls the spread of the distribution.
For our task, we assume that the movement patterns of the radiologists' gaze are different in saccade and fixation periods, implying various values of $\gamma$ to derive, i.e., $\gamma_s$ for eye movement of ``seeking" (saccade points) and $\gamma_f$ for ``focusing" (fixation points). 

An example is illustrated in Fig. \ref{fig:denoise-comparison}(a), where the first row shows all recorded gaze points in a collected gaze track starting at 0s and ending at 10.03s. As we can compute the step length between any two consecutive gaze points in the track, we further fit their distribution to the model in Eq. (\ref{levy-flight}) and derive $\gamma=0.95$. However, if closely examining the gaze points in a specific time window of [0.22s, 0.89s] (cf. the second row in Fig. \ref{fig:denoise-comparison}(a)), one may derive $\gamma=0.63$ for the gaze points in the time window. The $\gamma$ value is significantly lower than a second time window of [6.44s, 7.11s] ($\gamma=1.67$) as in the last row of the figure. Note that the two time windows correspond to examples of the saccade and fixation periods, respectively.

When the vision system is paying more attention, the step length of the gaze tends to be more compactly distributed~\cite{brockmann1999human}, implying $\gamma_f<\gamma_s$. 
Using this rule, we can identify the fixation points where $\gamma$ is high within a sliding time window. Empirically, we use the sliding window of 0.67 second (60 points) in width for this purpose. Since all healthy images have no lesions, the scan paths are assumed to contain ``seeking" gaze only. 
We thus can use the healthy cases to calculate the threshold of attention level $\gamma_{th}$. Specifically, in our experiment, $\gamma_{th}$ is the mean of $\gamma$ from all the healthy cases. More discussions regarding this calculation are provided in supplementary materials. For a training image, if the calculated attention level $\gamma$ within a sliding window is smaller than $\gamma_{th}$, the gaze track in the sliding window is considered as fixation. 
On the contrary, if $\gamma > \gamma_{th}$, the sliding window captures saccade gaze points. Then, after calculating attention level $\gamma$ for every sliding window along the entire gaze track, only the gaze points in the fixation sliding windows ($\gamma>\gamma_{th}$) are preserved while other gaze points are discarded. 

\rv{In our experiments, we notice that the raw data from eye-tracker contains some information irrelevant to our task, such as covert attention and inter-observer variability. We further acquire the gaze attention map from the post-processed gaze points of each training image, by modeling the Gaussian mixtures with the kernel at the radius of $99$ pixels (with the sigma set by default to 30.2 as in OpenCV). This choice is highly related to the used monitor, image size and distance from eye to the screen. More discussion about this choice can be found in the supplemental materials.} 

We show 4 cases in Fig. \ref{fig:denoise-comparison}(b), including the original gaze points, the original attention maps, the post-processed gaze points and attention maps. 
One may notice that the post-processed gaze points and attention maps contain less abundant saccade noises. \rv{In summary, our post-processing algorithm can improve the localization performance of the gaze maps for lesion regions. For example, the intersection over the union (IoU:\%) between the detected region of the gaze maps and the manually labeled bounding boxes has increased from $31.9\pm13.3$ to $40.2\pm18.5$.} Detailed results can be found in supplementary Table S1.

\subsection{GA-Net: Gaze-Guide Attention Network}
The GA-Net, with its architecture shown in Fig. \ref{fig:framework}, aims to make the network attention being consistent with the external supervision of expert visual attention. 

To plainly demonstrate the validity of attention consistency, we adopt the commonly-used ResNet to build up the CAD system. 
With the backbone classification network, the proposed attention consistency (AC) module is added. 
Therefore, in the following, we will introduce our GA-Net from the perspectives of the classification backbone and the AC module, respectively.

\textbf{Classification Network.} 
We use ResNet-style~\cite{he2016deep} CNN as the backbone for our task (with details in Fig. \ref{fig:framework}), which takes pre-trained weights on ImageNet. We adopt dilated convolution on the last stage of ResNet and set the stride to 1 instead of 2. Thus, the output feature maps from the ResNet-style backbone have a size of $16\times16$. Then, we perform global average pooling, followed by a fully-connected layer to produce the desired classification output. We use cross-entropy loss between the network prediction and the ground-truth from radiologists as the classification loss.

\textbf{Class Activation Map.} We employ the CAM module~\cite{zhou2016learning} to probe the attention of the classification network. First, in Fig. \ref{fig:framework}, let $f_{k}(x,y)$ be the activation of the $k$th feature map in the last convolutional layer at the spatial location $(x,y)$. Then, for each class $c$, we have the prediction output $S^c$:
\begin{equation}
\label{eq:cls}
{S^c} = \sum\limits_k {(w_k^c \cdot \frac{1}{{X \cdot Y}}\sum\limits_{x,y} {{f_k}} (x,y))}  ,
\end{equation}
where $X,Y$ are the sizes of the feature maps. $w_k^c$ is the weight from the fully-connected layer for each class $c$, and $\frac{1}{{X \cdot Y}}\sum\limits_{x,y} {{f_k}} (x,y)$ represents the global average pooling operation within the $k$th feature map. So it can be observed that $w_k^c$ is the aggregation weight of the $K$ pooling feature vectors for the prediction output $S^c$ of each class $c$. Finally, after softmax operation, the classification probability can be derived from $S^c$. 

CAM is proposed to reveal the important regions for network decisions by propagating the weights of the fully-connected layer to the feature maps from the last convolutional layer. Specially, the neural network's attention map $A^c$ for the class $c$ can be formulated as 
\begin{equation}
\label{eq:cam}
   A^c = \sum_k w_k^c f_k(x,y).
\end{equation}

When comparing (\ref{eq:cls}) and (\ref{eq:cam}), it can be observed that CAM shares the weights from the fully-connected layer to aggregate the feature maps. 

In this paper, we take the CAM as an online training module to accommodate gaze supervision. 
As shown in Fig. \ref{fig:framework}, for each class, we share the weights $w$ (size: $1\times512$) from the fully-connected layer to multiply the corresponding feature maps.

\textbf{Attention Consistency with Homoscedastic Uncertainty.} 
Assuming the gaze data are ready to supervise the classification network, we now want to ensure the consistency of the network attention with the attention reflected by expert gaze tracking. To this end, we design the AC module and integrate it with the classification backbone. This AC module concerns the problem of optimizing the backbone with respect to giving more attention to the areas highlighted by the radiologist’s visual attention.

The naive approach to enforce this consistency could be simply minimizing the mean square error (MSE) between the two attention maps. Specifically, with $G$ for the gaze attention map and $A$ for the network attention map of the predicted class, the consistency loss in MSE can be represented by

\begin{equation}
\label{eq:mse}
L_{mse} = \sum_{x,y}(A(x,y)-G(x,y))^2,
\end{equation}
where $(x,y)$ refers to spatial location.

This MSE constraint strictly requires consistency between the network and gaze attention maps. However, in the real environment, there are always distortions even after denoising the gaze maps due to two reasons. 
First, the reading process always contains many random walks to seek for the disease lesions by quickly scanning the whole image. Second, different radiologists have different reading habits, which easily leads to different gaze maps even for the same medical image.

Since the radiologists typically focus on the lesions at most time, the variances of the gaze maps are usually small and finite. \rv{Therefore, inspired by \cite{kendall2018multi}, we propose to use the homoscedastic uncertainty to model visual attention and to suppress inter-observer variability.}
Specifically, we define the conditional Gaussian probabilistic model centered on the network attention map $A$:
\begin{equation}
\label{eq:guassian}
p(G|A) = \mathcal{N}(A,\sigma^2),
\end{equation}
where $\sigma$ quantifies the spread of the uncertainty. 
Then, the log-likelihood of the attention maps can be formulated as
\begin{equation}
\log (p(G|A)) \propto  - \frac{1}{{2{\sigma ^2}}}||G - {A}|{|^2} - {\rm{log}}\sigma .
\end{equation}

For the network optimization, we maximize this log-likelihood. 
Thus, by combining this new optimization goal with the MSE constraint, we propose the attention consistency loss $L_{ac}$ as:
\begin{equation}
\label{eq:AC_Loss}
L_{ac}= \frac{1}{2\sigma^2}\frac{1}{{X \cdot Y}}\sum_{xy}(A(x,y)-G(x,y))^2+\text{log}\sigma,
\end{equation}
where $\sigma$ is a learnable parameter to mitigate the hidden uncertainty in the gaze map $G$. Note that when $\sigma=0$, the loss in Eq. (\ref{eq:AC_Loss}) reduces to the MSE loss in Eq. (\ref{eq:mse}).

\subsection{Implementation Detail}
\rv{In all three sets of experiments, our network is pre-trained on ImageNet. We set the batch size to 32 and the initial learning rate to 0.001 for the Adam optimizer, and train the network for 30 epochs. All the experiments are implemented with PyTorch and run on a single NVIDIA Titan RTX GPU. More detailed settings can be found on our project website.}

\begin{figure*}[t]
    \centering
    \includegraphics[width=0.9\textwidth]{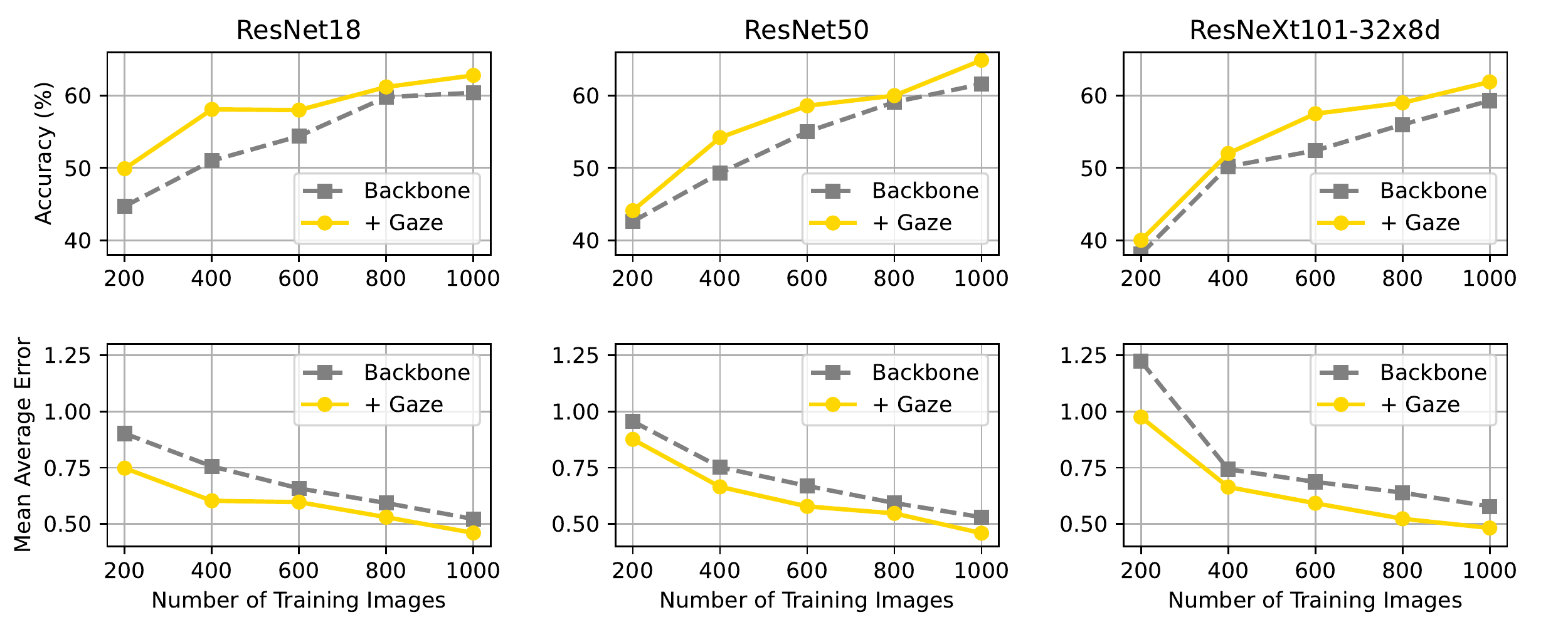}
    \caption{Comparisons of the classification performance in the test set, when the gaze attention is enabled or disabled. From left to right: the backbone of ResNet18, ResNet50, and ResNeXt101-32x8d.}
    \label{fig:knee-performance-plot}
\end{figure*}

\begin{table*}[h]
\caption{Quantitative classification result on the test set. Three different classification backbones, with the gaze attention enabled or disabled, are compared when different numbers of training images are available.}
\label{different-backbone}
\centering
\begin{tabular}{ccccccccccc}
\hline
\multirow{2}{*}{} & \multicolumn{2}{c}{200 Training Images} & \multicolumn{2}{c}{400 Training Images} & \multicolumn{2}{c}{600 Training Images} & \multicolumn{2}{c}{800 Training Images} & \multicolumn{2}{c}{1000 Training Images} \\ \cline{2-11} 
                        & ACC           & MAE             & ACC           & MAE             & ACC           & MAE             & ACC           & MAE             & ACC            & MAE             \\ \hline
ResNet-18               & 0.447              & 0.902           & 0.510               & 0.756           & 0.544              & 0.658           & 0.598              & 0.593           & 0.604               & 0.522           \\ \hline
ResNet-18+Gaze           & \textbf{0.499}             & \textbf{0.748}           & \textbf{0.581}              & \textbf{0.603}           & 0.580              & 0.597           & \textbf{0.612}              & \textbf{0.530}            & 0.628               & 0.460            \\ \hline
ResNet-50               & 0.426              & 0.956           & 0.493              & 0.752           & 0.550              & 0.669           & 0.591              & 0.593           & 0.616               & 0.530            \\ \hline
ResNet-50+Gaze          & 0.441              & 0.876           & 0.542              & 0.665           & \textbf{0.586}              & \textbf{0.578}           & 0.600              & 0.547           & \textbf{0.649}               & \textbf{0.459}           \\ \hline
ResNeXt-101-32x8d       & 0.381              & 1.223           & 0.502              & 0.743           & 0.524              & 0.687           & 0.560              & 0.639           & 0.593               & 0.578           \\ \hline
ResNeXt-101-32x8d+Gaze  & 0.400              & 0.975           & 0.520              & 0.664           & 0.575              & 0.592           & 0.590              & 0.523           & 0.619               & 0.482           \\ \hline
\end{tabular}
\end{table*}

\section{Experimental Result}
Our experiments are three-fold. First, we demonstrate that, by adding visual attention (i.e., gaze), the classification performance can benefit several different classification backbones. Second, we compare our GA-Net with state-of-the-art methods such as popular self-learned attention, to validate the effectiveness of utilizing additional expert visual attention for supervision. Third, we compare our visual attention generated from gaze with the attention generated by manually-drawn bounding boxes, to verify whether eye-tracking is more efficient than the bounding-box annotation for external supervision.

\begin{figure*}[ht]
    \centering
    \includegraphics[width=0.96\textwidth]{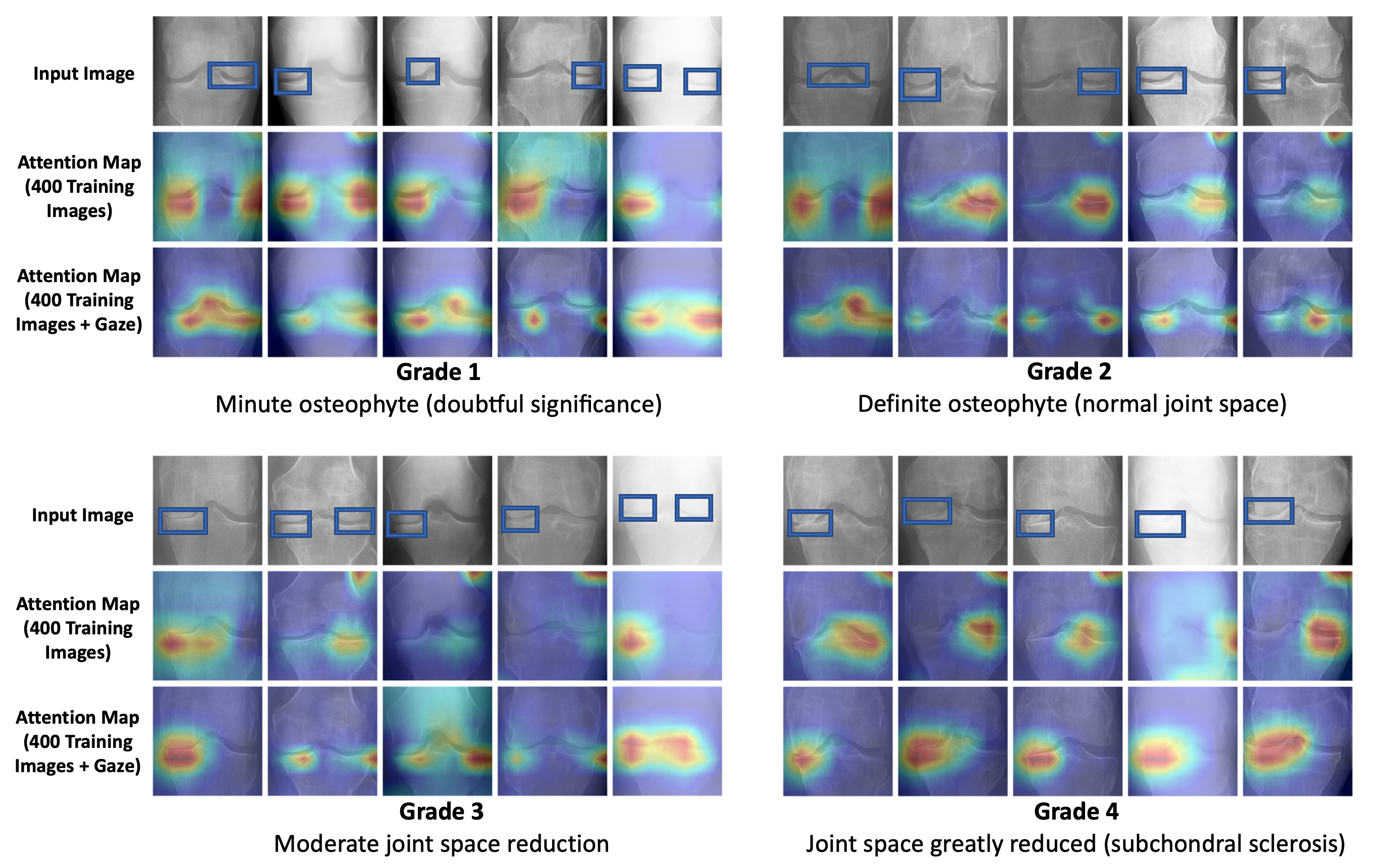}
    \caption{Comparison of network attention maps on the test set. The first row in each grade shows typical input images, with bounding boxes to highlight the abnormality related to OA. The second row shows the attention maps generated by the ResNet-50 trained on 400 images. The third row, which uses gaze information and the same number of images for training, shows better abnormality localization than the second row. }
    \label{fig:camvis}
\end{figure*}

\subsection{Experimental Setup}

\textbf{Image Data.} Knee X-ray images used for this study are obtained from the osteoarthritis initiative (OAI), which is a multi-center, longitudinal study targeting OA\cite{nevitt2006osteoarthritis}. The KL grading system\cite{kellgren1957radiological} is the most commonly used knee OA severity grading system, which ranks knee OA severity from Grade 0 (i.e. normal) to Grade 4. \rv{We randomly select 1000 training images, 200 validation images and 1000 test images from OAI X-ray radiograph repository which has been made available by Chen et al. \cite{chen2019fully}. The above three image sets are exclusive in terms of patients. During the training, there could be a single image (i.e., left knee or right knee) or two images (i.e., left knee and right knee) per case, which are treated independently.} After the splitting, the test set contains 385 images of Grade 0, 179 images of Grade 1, 270 images of Grade 2, 135 images of Grade 3, and 31 images of Grade 4. 

\textbf{Gaze Data.} The gaze data are collected from two experienced radiologists using the gaze collection paradigm previously mentioned in Section \ref{gaze collection and post-processing}. The gaze data collection is upon the training set, and completed in three times. 
In the first two experiments (Section \ref{contribution of gaze attention} and Section \ref{comparison with the state-of-the-art}), 100 gaze tracks are used, containing 29 images of Grade 1, 40 images of Grade 2, 21 images of Grade 3, and 10 images of Grade 4. In the last experiment  (Section \ref{comparison with manual bounding box}), we use 200 gaze tracks, containing 55 images of Grade 1, 81 images of Grade 2, 40 images of Grade 3, and 24 images of Grade 4.
In addition, we record 154 tracks for KL Grade 0, which are used for calculating $\gamma_{th}$. 

\textbf{Metrics.} As a classification task, we choose the accuracy (ACC) on the test set as a primary metric. Moreover, in clinical diagnosis, mis-classifying the KL grade of a case to its faraway grade (e.g., mis-classifying Grade 0 to Grade 4) is far more consequential than mis-classifying two nearby grades (e.g., from Grade 0 to Grade 1). \rv{Therefore, we also use the mean absolute error (MAE) between the predicted grade and the ground-truth grade (given by experts) as a complementary metric to evaluate the classification performance.} The two metrics are also adopted in early literature report~\cite{chen2019fully}.

\rv{
\textbf{Statistical Analyses.} To validate that our method delivers better classification performance, a t-test is performed for comparing the MAEs. For example, in Table I, we compare with the same network architecture, e.g. between ResNet18+Gaze and ResNet18, to test the hypothesis that having gaze yields better classification accuracy than without gaze. We also statistically compare the time consumption of two labeling types, i.e., placing the bounding boxes and collecting eye gaze. More details can be found in the supplementary materials.}

\begin{table*}[h]
\caption{Comparison of performance on training sets of five different sizes.}
\label{sota_table}
\centering
\begin{tabular}{ccccccccccc}
\hline
               & \multicolumn{2}{c}{200 Training Images}                           & \multicolumn{2}{c}{400 Training Images}                           & \multicolumn{2}{c}{600 Training Images}                           & \multicolumn{2}{c}{800 Training Images}                           & \multicolumn{2}{c}{1000 Training Images}                          \\ \cline{2-11} 
               & ACC                             & MAE                             & ACC                             & MAE                             & ACC                             & MAE                             & ACC                             & MAE                             & ACC                             & MAE                             \\ \hline
SE             & 0.447                           & 0.910                           & 0.530                           & 0.662                           & 0.536                           & 0.616                           & 0.583                           & 0.549                           & 0.582                           & 0.554                           \\ \hline
CBAM           & 0.431                           & 0.949                           & 0.470                           & 0.839                           & 0.495                           & 0.774                           & 0.518                           & 0.639                           & 0.544                           & 0.647                           \\ \hline
ResNeXt        & 0.445                           & 0.876                           & 0.497                           & 0.752                           & 0.538                           & 0.638                           & 0.557                           & 0.582                           & 0.574                           & 0.579                           \\ \hline
ViT-B/16       & 0.449                           & 0.832                           & 0.531                           & 0.674                           & 0.555                           & 0.621                           & 0.576                           & 0.545                           & 0.606                           & 0.526                           \\ \hline
ViT-L/16       & \textbf{0.492} & 0.819                           & 0.543                           & 0.620                           & 0.559                           & 0.584                           & \textbf{0.600} & 0.540                           & 0.607                           & 0.537                           \\ \hline
EffientNet-b5  & 0.454                           & 0.818                           & 0.521                           & 0.688                           & 0.562                           & 0.593                           & 0.572                           & 0.566                           & 0.596                           & 0.533                           \\ \hline
EffientNet-b6  & 0.460                           & 0.879                           & 0.539                           & 0.632                           & 0.551                           & 0.613                           & 0.587                           & 0.556                           & 0.583                           & 0.564                           \\ \hline
EffientNet-b7  & 0.483                           & 0.767                           & 0.546                           & 0.613                           & 0.556                           & 0.602                           & 0.588                           & 0.525                           & 0.599                           & 0.544                           \\ \hline
\rv{Dalia et al.} & 0.387                           & 1.223                           & 0.512                           & 0.711                           & 0.524                           & 0.687                           & 0.575                           & 0.592                           & 0.594                           & 0.570                           \\ \hline
\rv{Jain et al.-1}  & 0.424                           & 0.975                           & 0.498                           & 0.749                           & 0.554                           & 0.617                           & 0.597                           & 0.546                           & 0.601                           & 0.550                           \\ \hline
\rv{Jain et al.-2}  & 0.483                           & 0.767                           & 0.544                           & 0.666                           & 0.588                           & 0.577                           & 0.586                           & 0.603                           & 0.603                           & 0.516                           \\ \hline\hline
ResNet+Gaze    & 0.441                           & 0.876                           & 0.542                           & 0.620                           & \textbf{0.567} & \textbf{0.550} & 0.595                           & 0.515                           & 0.617                           & 0.487                           \\ \hline
SE+Gaze        & 0.480                           & \textbf{0.736} & \textbf{0.561} & \textbf{0.602} & 0.564                           & 0.567                           & \textbf{0.600} & \textbf{0.499} & \textbf{0.618} & \textbf{0.470} \\ \hline

\end{tabular}
\end{table*}
\subsection{Contribution of Gaze Attention}
\label{contribution of gaze attention}
We first compare the cases when the gaze attention is enabled (through attention consistency) or disabled in training the classification backbones. For comprehensive comparisons, we adopt three different classification backbones, each of which is widely applied in medical image analysis studies. The comparison is shown in Table \ref{different-backbone}. In particular, “ResNet-18” (the first row in the table) indicates that only the given numbers of training images are available to supervise the ResNet-18 classification network, while “ResNet-18+Gaze” means that 100 extra gaze attention maps are provided. The gaze attention maps are corresponding to 100 images in the training set, which are used in all comparisons in Table \ref{different-backbone}. One can observe from the table that, by comparing “ResNet-18” and “ResNet-18+Gaze”, the test accuracy has risen from 0.447 to 0.499, when only 200 images are available for training. Similar results can be found by comparing “ResNet-50” and “ResNet50+Gaze”, as well as “ResNeXt-101-32x8d” and “ResNeXt101-32x8d+Gaze”. The above results confirm the validity and effectiveness of adding radiologists’ visual attention, regardless of the specific type of the classification network.

Then, we investigate the contributions of the extra gaze attention maps, given increasing numbers of training images. As in Table \ref{different-backbone} and Fig. \ref{fig:knee-performance-plot}, we observe consistently increasing classification performance, when more training images are available, with or without additional supervision from the gaze attention. Meanwhile, the gain from adding the gaze attention is generally stable, regardless of a few or many training images provided (i.e., from 200 to 1000). \rv{In particular, the p-values are smaller than 0.01 when comparing under every training setting, between ``ResNet18'' and ``ResNet18+Gaze", ``ResNet18'' and ``ResNet18+Gaze", as well as ``ResNeXt101'' and ``ResNeXt101+Gaze". All t-test results can be found at Supplementary Table S2
.} The contributions of the gaze attention are evidently clear for all three investigated classification backbones.
\begin{figure}
    \centering
    \includegraphics[width=0.48\textwidth]{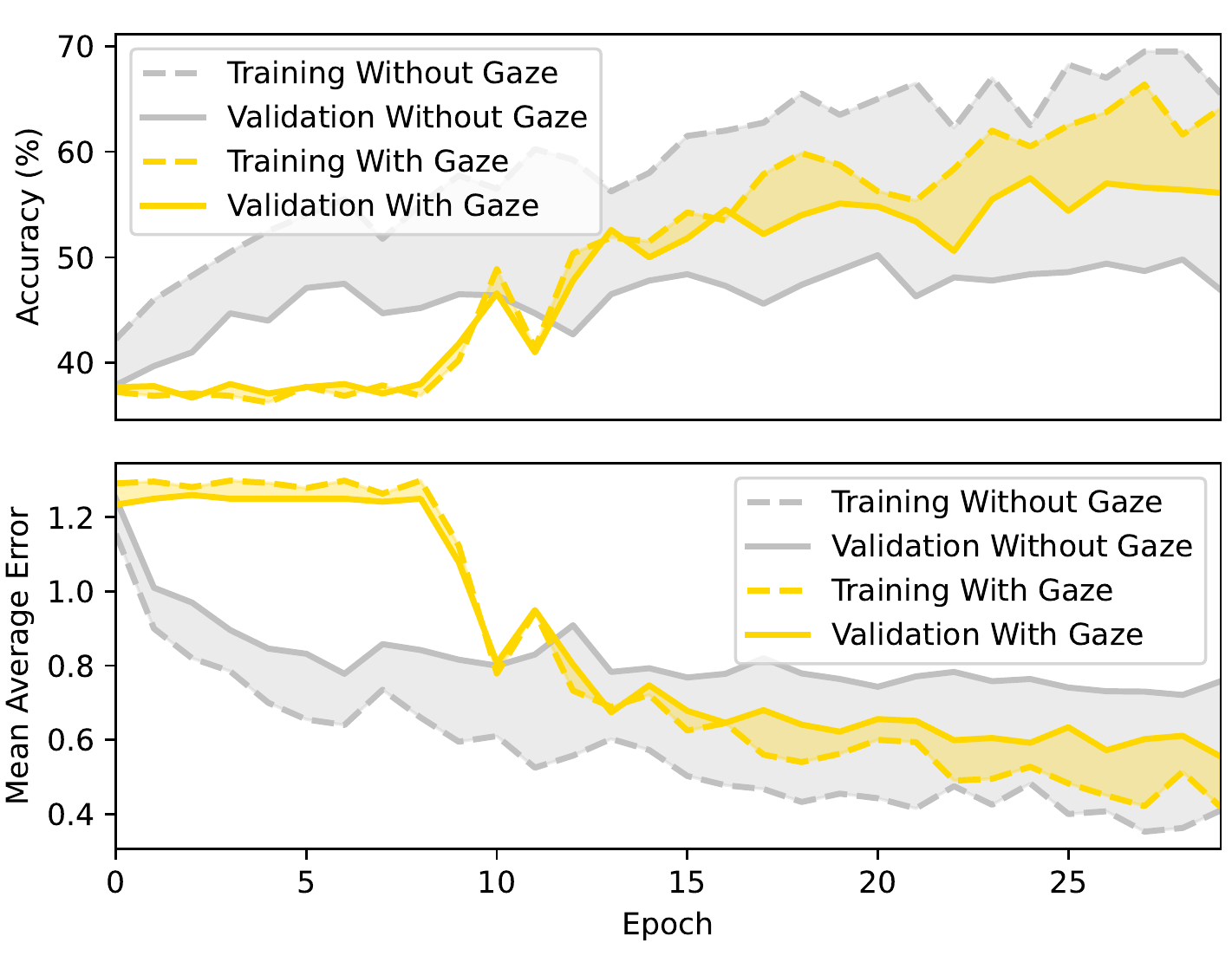}
    \caption{The classification metrics on the training and validation sets, when the training process is evolving with more epochs.}
    \label{fig:overfit}
\end{figure}
\begin{figure}
    \centering
    \includegraphics[width=0.48\textwidth]{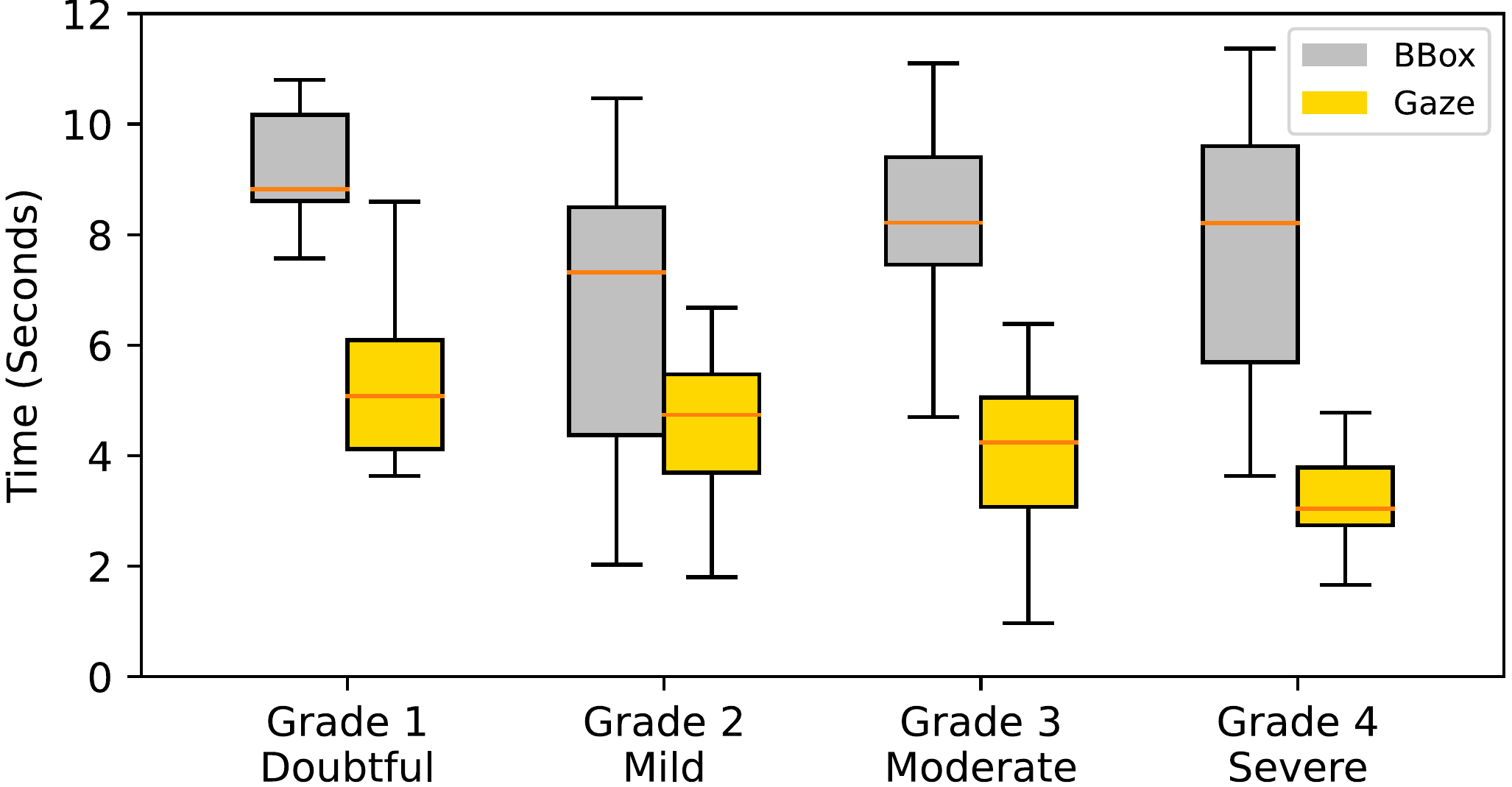}
    \caption{Time cost comparison between collecting bounding boxes and collecting gaze. On average, Grade 1 costs 45.5\% less time to collect, Grade 2 saves 30.9\%, Grade 3 saves 50.9\%, and Grade 4 saves 60.1\%.}
    \label{fig:time-comparison}
\end{figure}

We further probe the possible reasons that underline the performance gain with the gaze attention enabled in training the classification networks. Particularly, we use CAM to look for the spatial parts of the images that are visually salient for the network to reach the predictions. In Fig. 6, we show the CAM outputs of “ResNet50” and “ResNet-50+Gaze” with 400 training images. When adding the radiologist’s gaze for training, “ResNet-50+Gaze” focuses on the abnormality with better precision of spatial localization. On the contrary, without gaze supervision, CAMs are fuzzier with spilled coverage to many redundant parts in the images. This result proves that the radiologist's gaze can truly improve the network's ability to acquire better abnormality localization.

\begin{table*}[h!]
\caption{Performance comparison of using bounding box attention and gaze attention.}
\label{bbox_table}
\centering
\begin{tabular}{ccccccccc}
\hline
\multirow{2}{*}{} & \multicolumn{2}{c}{400 Training Images}  & \multicolumn{2}{c}{600 Training Images} & \multicolumn{2}{c}{800 Training Images}  & \multicolumn{2}{c}{1000 Training Images} \\ \cline{2-9} 
                        & ACC       & MAE            & ACC       & MAE           & ACC       & MAE            & ACC       & MAE            \\ \hline
ResNet-50               & 0.493          & 0.752          & 0.550          & 0.669         & 0.591          & 0.593          & 0.616          & 0.515          \\ \hline
+100 BBox                & \textbf{0.544} & 0.666          & 0.588          & 0.577         & 0.603          & 0.555          & 0.610           & 0.553          \\ \hline
+200 BBox                & 0.498          & 0.749          & \textbf{0.603} & \textbf{0.550} & \textbf{0.615} & 0.522 & 0.638          & 0.496          \\ \hline
+100 Gaze                  & 0.542          & \textbf{0.665} & 0.567          & \textbf{0.550}         & 0.595          & \textbf{0.515}          & \textbf{0.649}          & \textbf{0.459}         \\ \hline
+200 Gaze                  & 0.528          & 0.716          & 0.586          & 0.578         & 0.600          & 0.554          & 0.617 & 0.487 \\ \hline
\end{tabular}
\end{table*}

Fig. \ref{fig:overfit} compares the training process when the gaze attention is enabled or disabled. Particularly, we use the ResNet-50 backbone for classification, with 400 training images and 100 gaze attention maps. In the beginning epochs, the gaze attention improves the network slowly; but after 10 epochs, the supervision from the gaze helps gain significantly. The classification performance gap between the training set and the validation set (i.e., shadowed areas in Fig. \ref{fig:overfit}) is much narrower when the gaze attention is available. This indicates that the trained networks are less likely to overfit with gaze supervision enabled. This improvement is in line with our expectation that the guidance offered by gaze can prevent the network from simply ``remembering" the whole image especially when trained with \rv{a} small dataset.

\subsection{Comparison with the State-of-the-art Attention Mechanism}
\label{comparison with the state-of-the-art}
We further compare the gaze attention with state-of-the-art methods to demonstrate the superiority of integrating radiologist’s visual attention into CAD. We particularly compare to the popular learning-based attention models, namely the channel\rv{-}attention based squeeze-and-excitation net (SE)~\cite{hu2018squeeze}, block attention module ResNet (CBAM)~\cite{woo2018cbam} and Vision Transformer (ViT-B/16 and ViT-L/16)\cite{dosovitskiy2020image}. These works demonstrate that the attention mechanism works as a powerful plug-in in various computer vision tasks. 

The comparisons are shown in Table \ref{sota_table}. We can observe that ``ResNet+Gaze" and ``SE+Gaze" outperform other learning-based attention methods in most cases. Note that ``ResNet”, ``SE” and ``CBAM” (the first three rows in the table) are all set to have the same depth of 50 layers for a fair comparison. One may note that in Table \ref{different-backbone}, ResNet-18 shows better performance in the cases with 200, 400, and 800 training images, respectively. However, we choose ResNet-50 instead of ResNet-18 in Table \ref{sota_table}, because we cannot find available weights that are pre-trained from ImageNet for CBAM ResNet-18. In general, the above results confirm that the gaze attention from human experts is better to guide the learning of the classification network, especially when have limited training data.

We also compare our GA-Net with other more sophisticated models, i.e., ResNeXt-50-32x4d~\cite{xie2017aggregated} (ResNeXt) and state-of-the-art natural image classification model EfficientNet~\cite{tan2019efficientnet}. EfficientNet is a set of networks designed by the neural architecture search technique. In our comparison, we choose three best-performing settings: EfficientNet-b5, EfficientNet-b6 and EfficientNet-b7. \rv{Moreover, we also compared two latest works designed for knee images: DeepOA\cite{dalia2021deepoa} (Dalia et al.) and OsteoHRNet\cite{jain2021knee} (Jain et al.). For OsteoHRNet, we used the biggest and mid-size HRNet in our reimplementation, i.e., HRNet-W64-C (Jain et al.-1) and HRNet-W30-C (Jain et al.-2).} The results in Table \ref{sota_table} show that our proposed method can lead ahead of much more complex networks generated by network architecture searching. \rv{Specifically, the t-tests over MAEs show that our method (``SE+Gaze") performs significantly better than 11 methods in 5 training settings (except ``EfficientNet-b7'' with 400 training images and ``Jain et al.-2'' with 600 training images). The details for the above comparisons can be found in supplementary Table S3.} The results further validate the merit of using radiologist’s gaze for external supervision as attention.

\subsection{Comparison with Manual Bounding Box Attention}
\label{comparison with manual bounding box}
In our proposed method, the gaze serve\rv{s} as the role of fine-granularity annotation other than image-level annotation. As aforementioned, using segmentation masks and bounding boxes can also provide extra finer-granularity supervision\cite{ouyang2020learning,li2018tell}. In our task, the abnormality can be roughly located by drawing a bounding box, although the drawing process needs to be actively initiated by the radiologist and thus costs additional labor in diagnosis. To this end, we compare our method with the alternative scheme of injecting bounding-box-based supervision, by referring to both classification performance and time cost when annotating data.

\textbf{Collecting Time Cost.} We collect the bounding box annotations from the same radiologists, from whom we also collect gaze data. Radiologists locate the abnormality in the images by drawing bounding boxes in LabelMe\cite{russell2008labelme}. For every KL-Grade, they have been asked to label 50 images. Fig. \ref{fig:time-comparison} shows the box-plot of time cost in annotating the image. As shown in the figure, gaze collection is more time-saving than bounding boxes. From Grade 0 to Grade 4, collecting gaze saves more than 30\% in time than collecting bounding boxes. Specifically, when collecting gaze for Grade 4, it is 60.1\% faster, because radiologists sometimes have to draw multiple bounding boxes due to the severity of the disease in high grade. \rv{The p-values of t-tests are $5.32\times 10^{-35}$ for Grade 1, $2.55\times 10^{-14}$ for Grade 2, $1.00\times 10^{-8}$ for Grade 3 and $2.66\times 10^{-31}$ for Grade 4.} The above result shows that, as an extra annotation to deep learning, the collection of the gaze is economically efficient compared to the widely used bounding boxes.

\textbf{Classification Performance.} We utilize the bounding boxes following\cite{ouyang2020learning}. The bounding boxes are used to generate binary attention maps, which are then fed as the attention ground-truth for the AC module. In Table \ref{bbox_table}, the results show that our gaze supervising GA-Net (``+100 Gaze" and ``+200 Gaze") attains comparable performance to the bounding-box supervising model (``+100 BBox" and ``+200 BBox") with different numbers of training images. \rv{In Supplementary Table S4, using gaze shows better performance in three training settings, using bounding boxes shows better performance in two settings. In addition, the two annotation ways are not significantly different in three other settings (i.e., $p\geq 0.01$)} We thus conclude that our gaze attention and AC module can leverage gaze to deliver similar performance in classification accuracy, compared with the case of using the bounding-box annotation, yet at a lower cost when annotating data.

\section{Conclusion and Limitation}
In this paper, we explore and demonstrate the importance of using radiologists' gaze for an intelligent assessment of OA upon knee X-rays. We propose a framework for providing direct guidance on the attention map generated by a diagnosis deep network in order to teach the network where to look. And we have built a toolkit for collecting and processing radiologists' gaze to generate accurate gaze attention maps. Extensive experiments show that our proposed method confidently outperforms state-of-the-art methods with no need for extra annotation time. 

With the above innovations and advantages, there are still some limitations that can be improved in our future work.
First, we only demonstrate over knee X-ray images for OA, which is in 2D. Note that the abnormality pattern in X-ray images of knee OA is relatively simple (i.e., most abnormality locates near the lateral ends of the narrowed joint space). Thus, many additional works are needed, for adapting our method to the scenarios of 3D modalities (e.g., MRI and CT), various diseases, and abundant clinical applications.
Second, we have assumed a plain uncertainty model, plus the denoising process, to handle the gaze tracks collected from two radiologists in this study. 
There is no doubt that the gaze is quite subjective and the variation could be high across individual experts. 
Thus, it is necessary to develop a closed-loop human-machine interaction scheme for improving diagnosis performance, i.e., via active learning.
Finally, there is also room to better utilize the gaze supervision. We have used a straightforward way to enforce the consistency between the gaze attention and the network attention. 
Our experimental results show that, although the proposed strategy is effective to deliver improved diagnosis performance, it is still an open question to optimally utilize the gaze, especially when having different numbers of training samples.

\rv{In the future, we plan to explore the usage of eye-tracking on more challenging modalities, e.g., mammography and CT images. And we plan to collect more behaviors from radiologists other than eye gaze, such as pupil size, voice, and mouse cursor movement.}

\bibliography{ref.bib}
\bibliographystyle{IEEEtran}
\end{document}